\input harvmac
\noblackbox
\def\tilde{\widetilde}
\newcount\figno
\figno=0
\def\fig#1#2#3{
\par\begingroup\parindent=0pt\leftskip=1cm\rightskip=1cm\parindent=0pt
\baselineskip=11pt
\global\advance\figno by 1
\midinsert
\epsfxsize=#3
\centerline{\epsfbox{#2}}
\vskip 12pt
\centerline{{\bf Figure \the\figno:} #1}\par
\endinsert\endgroup\par}
\def\figlabel#1{\xdef#1{\the\figno}}

\def\np#1#2#3{Nucl. Phys. {\bf B#1} (#2) #3}
\def\pl#1#2#3{Phys. Lett. {\bf B#1} (#2) #3}


\font\cmss=cmss10
\font\cmsss=cmss10 at 7pt
\def\rlx{\relax\leavevmode}
\def\inbar{\vrule height1.5ex width.4pt depth0pt}
\def\IC{\relax\,\hbox{$\inbar\kern-.3em{\rm C}$}}
\def\IN{\relax{\rm I\kern-.18em N}}
\def\IP{\relax{\rm I\kern-.18em P}}
\def\ZZ{\rlx\leavevmode\ifmmode\mathchoice{\hbox{\cmss Z\kern-.4em Z}}
 {\hbox{\cmss Z\kern-.4em Z}}{\lower.9pt\hbox{\cmsss Z\kern-.36em Z}}
 {\lower1.2pt\hbox{\cmsss Z\kern-.36em Z}}\else{\cmss Z\kern-.4em
 Z}\fi} 
\def\IZ{\relax\ifmmode\mathchoice
{\hbox{\cmss Z\kern-.4em Z}}{\hbox{\cmss Z\kern-.4em Z}}
{\lower.9pt\hbox{\cmsss Z\kern-.4em Z}}
{\lower1.2pt\hbox{\cmsss Z\kern-.4em Z}}\else{\cmss Z\kern-.4em
Z}\fi}

\def\narrowplus{\kern -.04truein + \kern -.03truein}
\def\narrowminus{- \kern -.04truein}
\def\narrowminussub{\kern -.02truein - \kern -.01truein}

\def\frac#1#2{{#1\over #2}}

\def\IZ{\relax\ifmmode\mathchoice
{\hbox{\cmss Z\kern-.4em Z}}{\hbox{\cmss Z\kern-.4em Z}}
{\lower.9pt\hbox{\cmsss Z\kern-.4em Z}}
{\lower1.2pt\hbox{\cmsss Z\kern-.4em Z}}\else{\cmss Z\kern-.4em
Z}\fi}
\def\IB{\relax{\rm I\kern-.18em B}}
\def\IC{{\relax\hbox{$\inbar\kern-.3em{\rm C}$}}}
\def\ID{\relax{\rm I\kern-.18em D}}
\def\IE{\relax{\rm I\kern-.18em E}}
\def\IF{\relax{\rm I\kern-.18em F}}
\def\IG{\relax\hbox{$\inbar\kern-.3em{\rm G}$}}
\def\IGa{\relax\hbox{${\rm I}\kern-.18em\Gamma$}}
\def\IH{\relax{\rm I\kern-.18em H}}
\def\II{\relax{\rm I\kern-.18em I}}
\def\IK{\relax{\rm I\kern-.18em K}}
\def\IP{\relax{\rm I\kern-.18em P}}

\font\cmss=cmss10 \font\cmsss=cmss10 at 7pt
\def\IR{\relax{\rm I\kern-.18em R}}

%

%
%
\def\eqnn#1{\xdef #1{(\secsym\the\meqno)}\writedef{#1\leftbracket#1}%
\global\advance\meqno by1\wrlabeL#1}
\def\eqna#1{\xdef #1##1{\hbox{$(\secsym\the\meqno##1)$}}
\writedef{#1\numbersign1\leftbracket#1{\numbersign1}}%
\global\advance\meqno by1\wrlabeL{#1$\{\}$}}
\def\eqn#1#2{\xdef #1{(\secsym\the\meqno)}\writedef{#1\leftbracket#1}%
\global\advance\meqno by1$$#2\eqno#1\eqlabeL#1$$}

\lref\rpol{J. Polchinski, ``TASI Lectures on D-Branes,''
hep-th/9611050\semi J. Polchinski, S. Chaudhuri and C. Johnson,
``Notes on D-Branes,'' hep-th/9602052. } 
\lref\rBFSS{T. Banks, W. Fischler, S. H. Shenker, and L. Susskind, ``M
Theory As A Matrix Model: A Conjecture,''  
hep-th/9610043, Phys. Rev. {\bf D55} (1997) 5112.}
\lref\rwtensor{E. Witten, ``Some Comments on String Dynamics,''
hep-th/9507121.}
\lref\rstensor{A. Strominger, ``Open P-Branes,'' hep-th/9512059,
\pl{383}{1996}{44}.} 
\lref\rsdecoupled{N. Seiberg, ``New Theories in Six Dimensions and
Matrix Description of M-theory on $T^5$ and $T^5/\IZ_2$,''
hep-th/9705221.}
\lref\rmoore{A. Losev, G. Moore, and S. Shatashvili, ``M \& m's ,''
hep-th/9707250.}
\lref\rbrunner{I. Brunner and A. Karch, ``Matrix Description of
M-theory on $T^6$,'' hep-th/9707259.}
\lref\rgilad{A. Hanany and G. Lifschytz, ``M(atrix) Theory on $T^6$
and a m(atrix) Theory Description of KK Monopoles,'' hep-th/9708037.}
\lref\rwati{W. Taylor IV, ``D-Brane Field Theory on Compact Space,''
hep-th/9611042, \pl{394}{1997}{283}.} 
\lref\rozali{M. Rozali, ``Matrix Theory and U-Duality
in Seven Dimensions,'' hep-th/9702136 \pl{400}{1997}{260}.}
\lref\brs{M. Berkooz, M. Rozali and N. Seiberg,  ``Matrix Description
of M theory on $T^4$ and $T^5$'', hep-th/9704089.}
\lref\rwitten{E. Witten, ``Bound States of Strings and P-branes'',
\np{460}{1996}{335}, hep-th/9510135.}
\lref\susskind{L. Susskind, ``Another Conjecture about M(atrix)
Theory,'' hep-th/9704080.}

\Title{\vbox{\hbox{hep-th/9710009}\hbox{IASSNS-HEP-97/108}}}
{Why is the Matrix Model Correct?}

\smallskip
\centerline{Nathan Seiberg\footnote{$^1$}{seiberg@sns.ias.edu}}
\medskip\centerline{\it School of Natural Sciences}
\centerline{\it Institute for Advanced Study}
\centerline{\it Princeton, NJ 08540, USA}

\vskip 1in

\noindent 
We consider the compactification of M theory on a light-like circle as
a limit of a compactification on a small spatial circle boosted by a
large amount.  Assuming that the compactification on a small spatial
circle is weakly coupled type IIA theory, we derive Susskind's
conjecture that M theory compactified on a light-like circle is given
by the finite $N$ version of the Matrix model of Banks, Fischler,
Shenker and Susskind.  This point of view provides a uniform
derivation of the Matrix model for M theory compactified on a
transverse torus $T^p$ for $p=0,...,5$ and clarifies the difficulties
for larger values of $p$.

\vskip 0.1in
\Date{10/97}

About a year ago Banks, Fischler, Shenker and Susskind (BFSS) \rBFSS\
proposed an amazingly simple conjecture relating M theory in the
infinite momentum frame to a certain quantum mechanical system.  The
extension to compactifications on tori $T^p$ for $p=1,...,5$ was
worked out in \refs{\rBFSS\rwati\rozali\brs -\rsdecoupled}.  This
proposal was based on the compactification of M theory on a spatial
circle of radius $R_s$ in a sector with momentum $P={N \over R_s}$
around that circle.  In the limit of small $R_s$ M theory becomes the
type IIA string theory and the lowest excitations in the sector with
momentum $P$ are $N$ D0-branes \rpol.  When the D0-brane velocities
are small and the string interactions are weak the D0-branes are
described \rwitten\ by the minimal supersymmetric Yang-Mills theory
with sixteen supercharges (SYM).  When the velocities or the string
coupling are not small, this minimal supersymmetric theory is
corrected by higher dimension operators.  The suggestion of BFSS was
that M theory in the infinite momentum frame in uncompactified space
is obtained by considering the minimal SYM quantum mechanical system
in the limit $N,R_s,P \rightarrow \infty$.  For compactification on
$T^p$ the proposal of \refs{\rBFSS,\rwati} is to consider Dp-branes,
which are described by SYM in $p+1$ dimensions, and again to truncate
to the minimal theory. This proposal raised a few questions:
\item{1.} Why is this proposal correct?
\item{2.} Why is the theory with small $R_s$ related to the theory
with large $R_s$?
\item{3.} More specifically, the minimal supersymmetric theory is
corrected by higher dimension operators, which are important when
$R_s$ and the velocities are not small.  Why is the extrapolation from
the minimal theory, which is valid at small $R_s$, correct for large
$R_s$?
\item{4.}  Furthermore, for $p\ge 4$ the minimal theory is not
renormalizable and hence it is ill defined.  Then, higher dimension
operators must be included in the description.  They reflect the fact
that the theory must be embedded in a larger theory with more degrees
of freedom.  This theory for $p=4,5$ was found in
\refs{\rozali-\rsdecoupled}.  The procedure to find these extensions
of the minimal theories did not appear systematic.  What is then the
rule to construct the theory in different backgrounds?

Susskind noted that the finite $N$ Matrix model enjoys some of the
properties expected to hold only in the large $N$ limit and suggested
that it is also physically meaningful \susskind.  He suggested that
the matrix model describes M theory compactified on a {\it light-like}
circle of radius $R$ with momentum $P^+={N\over R}$.  Such a
compactification on a light-like circle with finite momentum is known
as the discrete light-cone and the quantization of this theory is
known as the discrete light-cone quantization (DLCQ).  With a
light-like circle the value of $R$ can be changed by a boost.
Therefore, the uncompactified theory cannot be obtained by simply
taking $R$ to infinity.  Instead, it is obtained by taking $R,N
\rightarrow \infty$ holding $P^+$ fixed.

In this note we will relate these two approaches to the Matrix theory.
In the process of doing so, we will derive the Matrix model and will
answer the questions above.  We will also present a uniform derivation
of the Matrix model for M theory on a compactified transverse space.

We start by reviewing some trivial facts about relativistic
kinematics.  A compactification on a light-like circle corresponds to
the identification
\eqn\llcir{{x \choose t} \sim {x \choose t} +{ {R\over\sqrt 2} \choose
- {R \over \sqrt 2}},}
where $x$ is a spatial coordinate, e.g. $x^{10}$.
We consider it as the limit of a compactification on a space-like
circle which is almost light-like
\eqn\allcir{{x \choose t} \sim {x \choose t} +{ \sqrt{ {R^2\over 2} +
R_s^2} \choose - {R\over \sqrt{2}} } \approx  {x \choose t} +{ {R
\over \sqrt{2}}  + {R_s^2 \over \sqrt{2} R} \choose - {R \over
\sqrt{2}}}} 
with $R_s \ll R$.  The light-like circle \llcir\ is obtained from
\allcir\ as $R_s \rightarrow 0$.  This compactification is related
by a large boost with 
\eqn\boostp{\beta = {R \over \sqrt{R^2+2 R_s^2}}\approx 1-
{R_s^2 \over R^2}}
to a spatial compactification on
\eqn\cpcir{{x \choose t} \sim {x \choose t} +{R_s \choose 0}.}

A longitudinal boost of the light-like circle \llcir\ rescales the
value of $R$.  It also rescales the value of the light-cone energy
$P^-$.  Therefore $P^-$ is proportional to $R$.  For small $R_s$ the
value of $P^-$ in the system with the almost light-like circle
\allcir\ is also proportional to $R$ (an exception to that occurs when
$P^-=0$ for the light-like circle; then $P^-$ can be non-zero for the
almost light-like circle).  The boost \boostp\ rescales $P^-$ to be
independent of $R$ and of order $R_s$ (if originally $P^-=0$, the
resulting $P^-$ after the boost can be smaller than order $R_s$).

Following
\ref\helpol{S. Hellerman and J. Polchinski, work in progress.}
(as referred to in
\ref\bbpt{K. Becker, M. Becker, J. Polchinski and A. Tseytlin,
``Higher Order Graviton Scattering in M(atrix) Theory,''
Phys.Rev. {\bf D56} (1997), 3174, hep-th/9706072.})
we now consider M theory compactified on a light-like circle \llcir\
as the $R_s \rightarrow 0$ limit of the compactification on an almost
light-like circle \allcir\ or as the limit of the boosted circle
\cpcir.  This way the DLCQ of M theory discussed in \susskind\ is
related to the compactification on a small spatial circle as in
\rBFSS.  For small $R_s$ the theory compactified on \cpcir\ is weakly
coupled string theory with string coupling $g_s=(R_s M_P)^{3\over 2}$,
and string scale $M_s^2=R_s M_P^3$ ($M_P$ is the Planck mass).  For
fixed energies and fixed $M_P$, the limit $R_s \rightarrow 0$ yields a
complicated theory with vanishing string scale.

However, as we said above, starting with $P^-$ of order one, the
effect of the boost is to reduce $P^-$ to be of order $R_s M_P^2$
($M_P^2$ is inserted on dimensional grounds).  This is exactly the
range of energies in the discussion of
\ref\dkps{M. Douglas, D. Kabat, P. Pouliot, and S. Shenker, 
``D-branes and Short Distances in String Theory'',
Nucl. Phys. {\bf B485} (1997) 85, hep-th/9608024.},
which was one of the motivations for the Matrix model of BFSS \rBFSS.
In order to focus on the modes with such values of $P^-$ we rescale
the parameters of the theory.  We do that by replacing the original M
theory, which is compactified on a light-like circle of radius $R$ by
another M theory, referred to as $\tilde M$ theory with Planck scale
$\tilde M_P$ compactified on a spatial circle of radius $R_s$.  The
transverse geometry of the original M theory is replaced by that of
the $\tilde M$ theory.  For example, for a compactification on a
transverse torus with radii $R_i$ the other theory has radii $\tilde
R_i$.

The relations between the parameters of these two theories are
obtained by combining the limit $R_s \rightarrow 0$ with $\tilde
M_P \rightarrow \infty$ holding $P^- \sim R_s \tilde M_P^2$ fixed.
Therefore we identify
\eqn\rslimi{R_s\tilde M_P^2 = R M_P^2,}
which is finite in the limit.  Since the boost does not affect the
transverse directions, we identify
\eqn\trans{M_PR_i = \tilde M_P \tilde R_i}
and keep it fixed.  In this limit the energies are finite and we
find string theory with string coupling and string scale
\eqn\scsc{\eqalign{
&\tilde g_s=(R_s \tilde M_P)^{3\over 2}= R_s^{3 \over 4}(RM_P^2)^{3
\over 4}\cr 
&\tilde M_s^2=R_s \tilde M_P^3=R_s^{-{1 \over 2}} (RM_P^2)^{3 \over
2}\cr .}} 
For $R_s \rightarrow 0$ with finite $M_P$ and $R$ we recover weakly
coupled string theory with large string tension.  This theory is very
simple and is at the root of the simplification of the Matrix model.

A sector with $P^+ = {N \over R} $ in the original M theory is mapped
to a sector of momentum $P= {N \over R_s} $ in the new $\tilde M$
theory.  In terms of this latter theory it includes $N$ D0-branes.
Therefore, the original M theory is mapped to the theory of D0-branes.
These D0-branes move in a small transverse space of size $\tilde R_i
\sim R_s ^{1 \over 2} \rightarrow 0$ (it is small even relative to the
string length $\tilde R_i \tilde M_s \sim R_s^{1 \over 4} \rightarrow
0$).

We conclude that M theory with Planck scale $M_P$ compactified on a
light-like circle of radius $R$ and momentum $P^+={N \over R}$ is the
same as $\tilde M$ theory with 
Planck scale $\tilde M_P$ compactified on a spatial circle \cpcir\
of radius $R_s$ with $N$ D0-branes in the limit
\eqn\limit{\eqalign{
&R_s \rightarrow 0 \cr
&\tilde M_P \rightarrow \infty \cr
&R_s\tilde M_P^2 = R M_P^2 = {\rm fixed} \cr
&\tilde M_P \tilde R_i =  M_P  R_i = {\rm fixed}. \cr}}
Here $R_i$ should be understood as generic parameters in the
transverse metric -- not only radii in a toroidal compactification.
Clearly, the general discussion applies to curved space with any
number of unbroken supercharges.

For a compactification on $T^p$ we can use T duality to map the
system of $N$ D0-branes to $N$ D$p$-branes on a torus with larger
radii
\eqn\sigmai{\Sigma_i = {1 \over \tilde R_i \tilde M_s^2} ={1 \over R_i
R M_P^3}.}
Note that $\Sigma_i$ are finite when $R_s \rightarrow 0$.  The string
coupling after this T duality transformation is 
\eqn\gsafd{\tilde g_s^\prime = \tilde g_s\tilde M_s^p \prod \Sigma_i=
\tilde M_s^{p-3} R^3 M_P^6 \prod \Sigma_i.} 
The low energy dynamics of these $N$ D$p$-branes are controlled by
$p+1$ dimensional SYM \rwitten\ with gauge coupling
\eqn\ymcoup{ g_{YM}^2 = {\tilde g_s^\prime \over \tilde M_s^{p-3}} = 
R^3 M_P^6 \prod \Sigma_i,}
which also has a finite $R_s \rightarrow 0$ limit.  Even though the
low energy dynamics of these D$p$-branes is finite in this limit, we
should explore the behavior at higher energies.  We will do that
shortly. 

To summarize, we have mapped the original M theory problem with a
light-like circle of radius $R$ and parameters $M_P$ and $R_i$ in the
sector with $P^+={N \over R}$ to a problem of $N$ D$p$-branes wrapping
a torus in string theory.  The radii of the torus, the string
coupling and string scale are
\eqn\nscss{\eqalign{
&\Sigma_i ={1 \over R_i R M_P^3}\cr 
&\tilde g_s^\prime= \tilde M_s^{p-3} R^3 M_P^6 \prod \Sigma_i \cr
&\tilde M_s^2=R_s^{-{1 \over 2}} (RM_P^2)^{3 \over 2} \cr}}
and $R_s \rightarrow 0$.  Exactly this limit was analyzed recently in
\ref\malda{J.M. Maldacena, ``Branes Probing Black Holes,''
hep-th/9709099.}.

Let us analyze this limit for various values of $p$.  For $p=0$ the T
duality which we performed is not necessary.  The theory is that of
D0-branes with vanishing string coupling and infinite string tension.
Since the gauge coupling $g_{YM}$ is finite, the theory is not
trivial.  The relevant degrees of freedom are strings stretched
between the D0-branes.  The infinite string scale decouples all the
oscillators on the strings.  Therefore the full theory is the minimal
SYM theory.  Note that closed strings or gravitons in the bulk of
space time decouple both because the string scale becomes large and
because the string coupling vanishes.  This is exactly the finite $N$
version of the Matrix model of BFSS \rBFSS.

For $p=1,2,3$ we recover the SYM prescription of \refs{\rBFSS,
\rwati}.  Again, the infinite string scale decouples the oscillators
on the strings which are stretched between the $N$ D$p$-branes.
Therefore the Lagrangian is that of the minimal SYM theory without
higher order corrections.  For $p=3$ the string coupling $\tilde
g_s^\prime$ does not vanish, but there are still no higher dimension
operators in the $3+1$ dimensional SYM, since they are all suppressed
by inverse powers of the string scale $\tilde M_s$.

For $p=4$ several new complications arise.  First, the low energy SYM
theory is not renormalizable and therefore cannot give a complete
description of the theory.  It breaks down at energies of order $1
\over g_{YM}^2$, where new degrees of freedom must be added.  Second,
the string coupling also diverges in our limit.  Therefore, in order
to analyze the system we need to study the strong coupling limit of
the $\tilde M$ theory, which is an eleven dimensional theory.  In this
limit the D4-branes become 5-branes wrapping the eleventh dimension.
Using
\nscss\ we find that this eleven dimensional theory is compactified on
a circle of finite radius
\eqn\newcir{\Sigma_5 = {\tilde g_s^\prime \over \tilde M_s} = R^3
M_P^6 \prod \Sigma_i,}
but its eleven dimensional Planck scale diverges
\eqn\newele{{\tilde M_s \over (\tilde g_s^\prime)^{1 \over 3}} \sim
R_s^{-{1 \over 6}} \rightarrow \infty.}
Since the eleven dimensional Planck scale is infinite, the modes in
the bulk of space-time decouple, and the theory on the brane is a 5+1
dimensional theory.  This non-trivial theory, known as the $(2,0)$
field theory, was first found in \refs{\rwtensor, \rstensor}.  The new
degrees of freedom, which we had to add at energy of order $1 \over
g_{YM}^2$, can now be interpreted as associated with momentum modes
around the circle \newcir.  These are related to instantons in the SYM
theory, which are D0-branes in the $\tilde M$ theory.  We have thus
derived the proposal of \refs{\rozali, \brs} to use this theory as a
Matrix theory for M theory on $T^4$.

A similar analysis applies to $p=5$.  Here we study the strong
coupling limit of D5-branes in IIB string theory.  Using S duality of
this theory we map it to NS5-branes in weakly coupled IIB theory.  We
thus recover the proposal of \refs{\brs,\rsdecoupled} for the
description of M theory on $T^5$ in terms of a new theory obtained by
studying NS5-branes in type II theory.  This theory, which can be
called a non-critical string theory, is not a local quantum field
theory.  In addition to the five sides of the torus $\Sigma_i$
\sigmai\ it is characterized by the string slope $\alpha^\prime =
g_{YM}^2 = R^3 M_P^6 \prod \Sigma_i$.  The SYM description breaks down
at energies of order $1 \over g_{YM}$, where new degrees of freedom
are added.  These degrees of freedom are the strings in the theory.
In terms of the $\tilde M$ theory and its type IIB string theory,
these are D1-branes.

For $p=6$ the situation is more complicated 
\ref\ganor{O. Ganor, ``On the M(atrix) Model for M Theory on $T^6$,''
hep-th/9709139.}.
Here we are led to consider the strong coupling limit of
D6-branes\foot{We thank E. Witten for very helpful discussions on this
limit.}.  As for $p=4$, this limit is described by an eleven
dimensional theory.  However, here the eleven dimensional Planck
scale
\eqn\plscs{ {\tilde M_s \over (\tilde g_s^\prime)^{1 \over 3}} = {1
\over RM_P^2 (\prod \Sigma_i)^{1 \over 3} }}
remains finite, but the radius of the eleventh dimension diverges
\eqn\newcirs{ {\tilde g_s^\prime \over \tilde M_s} \sim R_s^{-{1 \over
2}} \rightarrow \infty.}  
Since the radius diverges, the D6-branes, which are Kaluza-Klein
monopoles associated with the eleventh dimension, expand and become an
$A_{N-1}$ singularity.  The gauge coupling of the associated SYM
theory is given by the eleven dimensional Planck scale, which remains
finite.  Since the eleven dimensional Planck scale is finite, there is
no reason to assume that the gravitons in the bulk of the ALE space
with an $A_{N-1}$ singularity decouple from it (see the discussion in
\refs{\rmoore\rbrunner-\rgilad}).  From the $\tilde M$ theory point of
view (before the T duality transformation) these graviton modes can be
identified as Kaluza-Klein monopoles\foot{This interpretation is due
to L. Susskind.}, which wrap the small $T^6$.  Their energy is of
order $R_s^2$.  After the boost \boostp\ their energy is of order
$R_s$ and it vanishes as $R_s \rightarrow 0$.  Therefore, the DLCQ
theory has an infinite number of new massless modes.  

We conclude that M theory on $T^6$ and a light-like circle with
momentum $P^+={N\over R}$ is the same as M theory with Planck scale
\plscs\ compactified on $T^6$ with an $A_{N-1}$ singularity in the
non-compact spatial directions.  The eleven dimensional gravitons
propagate in the entire space.  There is also an $SU(N)$ SYM in 6+1
dimensions describing the interactions of gluons at the singularity.
Comparing with the situation for lower values of $p$, we seem to miss
a decoupled 6+1 dimensional $U(1)$ multiplet.  However, considering
the limit which leads to this configuration carefully, we see that the
$U(1)$ multiplet exists.  It is ``smeared'' over the four dimensional
non-compact space.

Unfortunately, this result is not satisfying.  The Matrix theory
offered a simple description of M theory, which can lead to useful
computations.  Here we see that for $p=6$, it goes over to a situation
which is apparently as complicated as the underlying M theory.

After the completion of this work we received a paper
\ref\asen{A. Sen, ``D0-Branes on $T^n$ and Matrix Theory,''
hep-th/9709220.} 
which partially overlaps with ours.  We also learned that
J. Polchinski had independently reached some of these conclusions.

\bigbreak\bigskip\bigskip\centerline{{\bf Acknowledgements}}\nobreak

We have benefited from discussions with O. Aharony, T. Banks,
J. Schwarz, S. Shenker, L. Susskind and E. Witten. This work is
supported in part by \#DE-FG02-90ER40542.

\listrefs

\end